\documentclass[aps,prb,twocolumn,superscriptaddress,floatfix]{revtex4-1}

\usepackage{graphicx}
\usepackage{dcolumn}
\usepackage{bm}
\usepackage[hidelinks]{hyperref}
\usepackage{siunitx}
\usepackage[version=4]{mhchem}
\usepackage{xcolor}
\usepackage{lipsum}

\DeclareSIUnit\sq{\ensuremath{\Box}}

\begin{document}

\preprint{xxxxxx}

\title{A Cryogenic Interface for Controlling Many Qubits}
  \author{S.~J.~Pauka$^*$}
    \affiliation{ARC Centre of Excellence for Engineered Quantum Systems, School of Physics, The University of Sydney, Sydney, NSW 2006, Australia.}
      \affiliation{Microsoft Quantum Sydney, The University of Sydney, Sydney, NSW 2006, Australia.}
\author{K.~Das$^*$}
    \affiliation{Microsoft Quantum Sydney, The University of Sydney, Sydney, NSW 2006, Australia.}
\author{R.~Kalra}
    \affiliation{Microsoft Quantum Sydney, The University of Sydney, Sydney, NSW 2006, Australia.}
    \author{A.~Moini}
    \affiliation{Microsoft Quantum Sydney, The University of Sydney, Sydney, NSW 2006, Australia.}
    \author{Y.~Yang}
  \affiliation{ARC Centre of Excellence for Engineered Quantum Systems, School of Physics, The University of Sydney, Sydney, NSW 2006, Australia.}
      \author{M.~Trainer}
    \affiliation{Microsoft Quantum Sydney, The University of Sydney, Sydney, NSW 2006, Australia.}
   \author{A.~Bousquet}
    \affiliation{Microsoft Quantum Sydney, The University of Sydney, Sydney, NSW 2006, Australia.}
  \author{C.~Cantaloube}
    \affiliation{Microsoft Quantum Sydney, The University of Sydney, Sydney, NSW 2006, Australia.}
  \author{N.~Dick}
    \affiliation{Microsoft Quantum Sydney, The University of Sydney, Sydney, NSW 2006, Australia.}
\author{G.~C.~Gardner}
    \affiliation{Birck Nanotechnology Center, Purdue University, West Lafayette, IN 47907, USA.}
    \affiliation{Microsoft Quantum Purdue, Purdue University, West Lafayette, IN 47907, USA.}
\author{M.~J.~Manfra}
    \affiliation{Department of Physics and Astronomy, Purdue University, West Lafayette, IN 47907, USA.}
    \affiliation{Birck Nanotechnology Center, Purdue University, West Lafayette, IN 47907, USA.}
    \affiliation{Microsoft Quantum Purdue, Purdue University, West Lafayette, IN 47907, USA.}
    \affiliation{School of Materials Engineering and School of Electrical and Computer Engineering, Purdue University, West Lafayette, IN 47907, USA.}
\author{D.~J.~Reilly$^{*\dagger}$}
        \affiliation{Microsoft Quantum Sydney, The University of Sydney, Sydney, NSW 2006, Australia.}
\affiliation{ARC Centre of Excellence for Engineered Quantum Systems, School of Physics, The University of Sydney, Sydney, NSW 2006, Australia.}
\date{\today}

\begin{abstract}
A scaled-up quantum computer will require a highly efficient control interface that autonomously manipulates and reads out large numbers of qubits, which for solid-state implementations are usually held at millikelvin (mK) temperatures. Advanced CMOS technology, tightly integrated with the quantum system, would be ideal for implementing such a control interface but is generally discounted on the basis of its power dissipation that leads to heating of the fragile qubits. Here, we demonstrate an ultra low power, CMOS-based quantum control platform that takes digital commands as input and generates many parallel qubit control signals. Realized using $\sim$ 100,000  transistors operating near 100 mK, our platform removes the need for separate control lines to every qubit by exploiting the low leakage of transistors at cryogenic temperatures to store charge on floating gate structures that are used to tune-up quantum devices. This charge can then be rapidly shuffled between on-chip capacitors to generate the fast voltage pulses required for dynamic qubit control. We benchmark this architecture on a quantum dot test device, showing that the control of thousands of gate electrodes is feasible within the cooling power of commercially available dilution refrigerators. 
\end{abstract}
\maketitle
\noindent{\bf Introduction}\\
{\em What technology can control a quantum computer?--} 
Executing a quantum application with potential for societal impact\cite{Reiher7555,Preskill2018quantumcomputingin,PhysRevLett.101.130504,Bravyi308,Aspuru-Guzik1704} will require a formidable number of high-fidelity qubits, operating in concert with a  control interface\cite{Dasxx,DJR_NPJ,PhysRevAppliedJohn,Lieven,McDermott,PhysRevApplied.3.024010,googleCMOS,geck2019control,DBLP:journals/jssc/PatraIDHSSSVBSC18} that passes signals between the classical- and quantum-domains of a quantum computer [see Fig. 1a]. At present, it is not explicitly known if, or how, this complex control interface can be realized at scale. Regardless of the details, it must address the significant input-output (IO) bottleneck that arises in quantum circuits due to  their inability\cite{nocloning} to fan-in and fan-out data in the way that classical processors operate\cite{Rent,EDM_Reilly}.   Rather, quantum computers have every logic-gate individually controlled by external signals,\cite{FRANKE20191,DJR_NPJ,lievenNPJ} bringing additional noise or heat load to the qubit system\cite{WallraffPRApp}. Brute-force management of these signals then presents a significant barrier to scale-up, a challenge that is well illustrated by a recent state-of-the-art experiment that uses $\sim$ 200 wideband coaxial-cables, 45 bulky microwave circulators, and racks of room temperature electronics to control just 53 qubits\cite{googlesup}.

Here, we report the realization of a chip-based cryogenic CMOS interface system that can be extended to control thousands of qubits. Without requiring the control system and qubits be monolithically integrated on the same substrate\cite{Zoo}, our architecture can leverage tight, chip-to-chip interconnects\cite{dense_packaging}  to manage the IO-bottleneck at the quantum-classical interface of semiconductor qubits based on majorana zero modes (MZMs)\cite{PhysRevB.95.235305}, electron spins,\cite{Petta_science05} or gatemon devices\cite{gatemon}.  Tight integration between qubits and their controllers is made possible by designing complex CMOS circuits that  function near 100 mK and with ultra low power dissipation. 
\begin{figure*}
    \includegraphics[width=\textwidth]{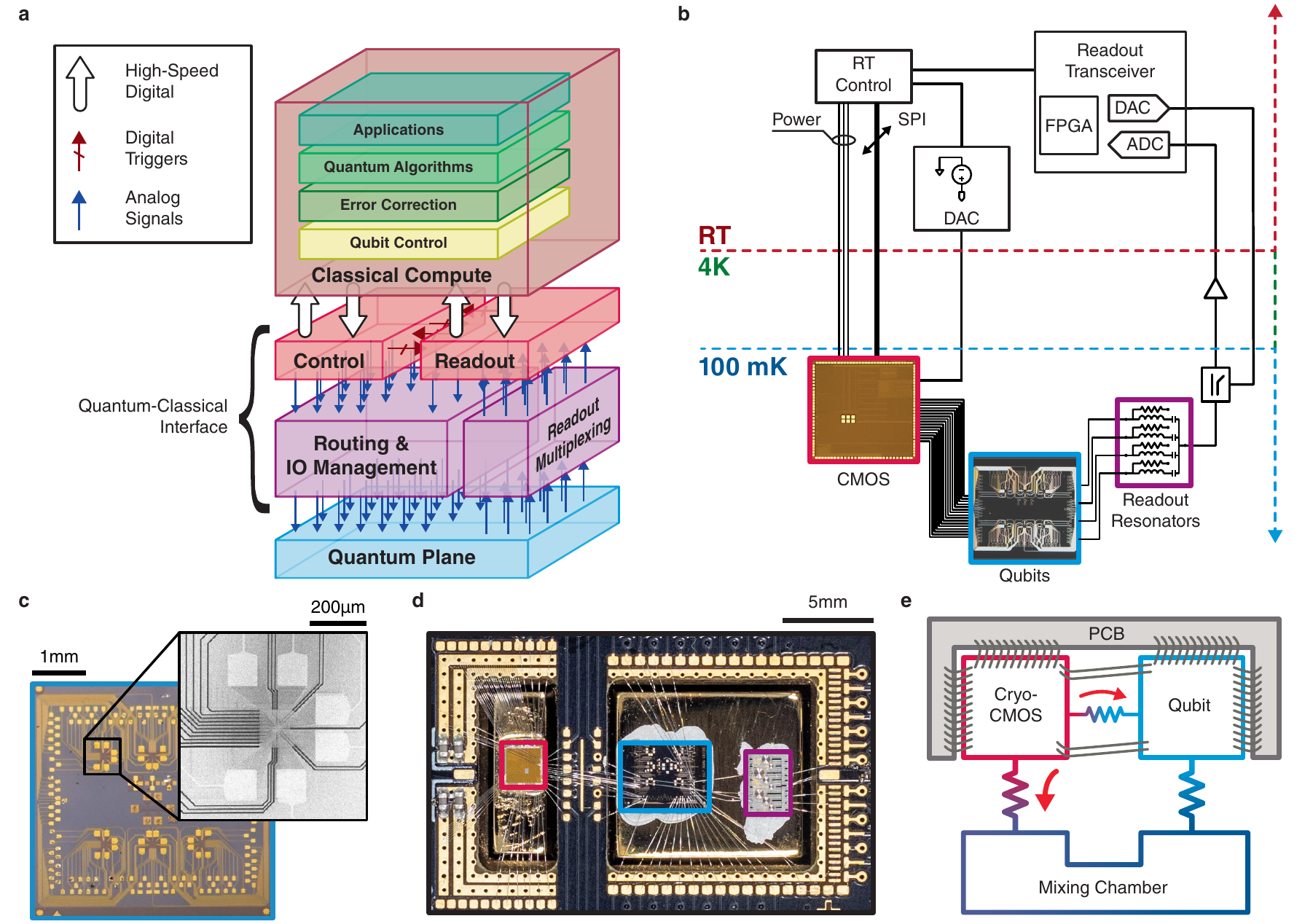}
    \caption{\label{fig:fig1} \textbf{The quantum-classical interface of a quantum computer.} \textbf{a.} The generic stack of elements needed for quantum computing. \textbf{b.} Control and readout sub-systems, distributed between room temperature and 100 mK. The brown cryo-CMOS chip addresses the IO bottleneck for control signals. \textbf{c.} Photograph and electron micrograph  of our qubit test platform based on 30 GaAs quantum dots [see Supp. Mat for details]. \textbf{d.} Photograph showing the cryo-CMOS chip (red box), qubit test chip (blue box), and resonator chip (purple box). Each chip is anchored to a gold-plated copper thermalization pillar, with a separate pillar used for the CMOS chip . \textbf{e.} Simplified thermal conductance model of the setup. The intended use of the partially separate cooling pillars is to increase the thermal conductivity to the mixing chamber (big red arrow) while reduce the direct heat (little red arrow) flowing from the hot CMOS chip to the qubit devices. 
     }
\end{figure*}
\begin{figure}
    \includegraphics[width=0.5\textwidth]{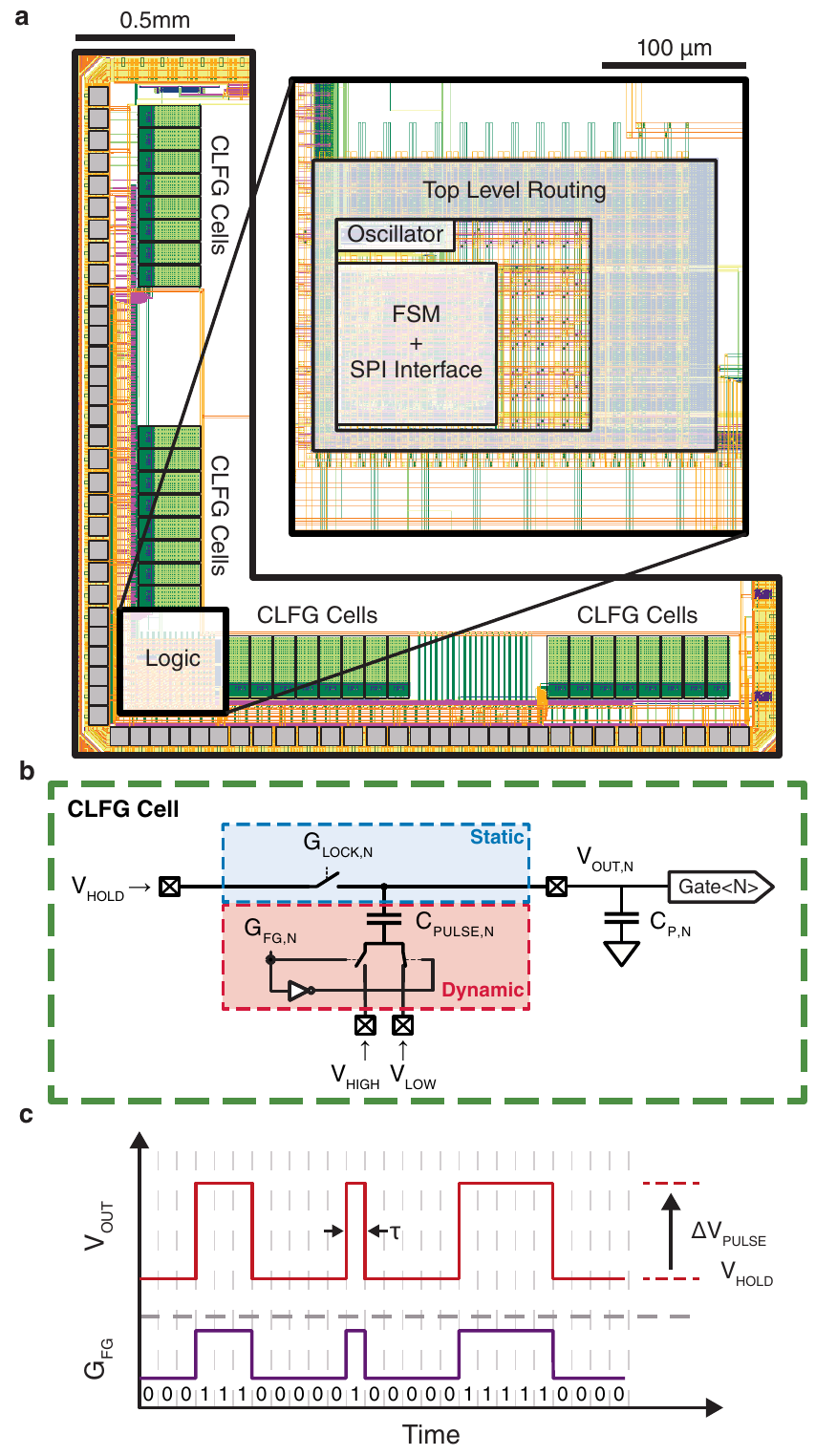}
    \caption{\label{fig:fig2} \textbf{Corner floor-plan and operation of the CMOS control-chip.} \textbf{a} The location of digital logic blocks (inset is a zoom of the block in the lower left corner). Tiled around the left and lower edge of the chip are 32 cells termed ``charge-lock fast-gate" (CLFG). These analog blocks generate static and dynamic voltages at their output (grey squares are bond pads). \textbf{b} A schematic of a single CLFG cell where $\textrm{C}_\textrm{P}$ is the sum of parasitic capacitances due to interconnect wiring on the CMOS and qubit chips. $\textrm{C}_\textrm{PULSE}$ is the on-chip capacitor. \textbf{c} Output voltage of the cell as $\textrm{G}_\textrm{FG,N}$ is pulsed, assuming a voltage $\textrm{V}_\textrm{HOLD}$ is locked on the cell.}
\end{figure}
\begin{figure*}
    \includegraphics[width=\textwidth]{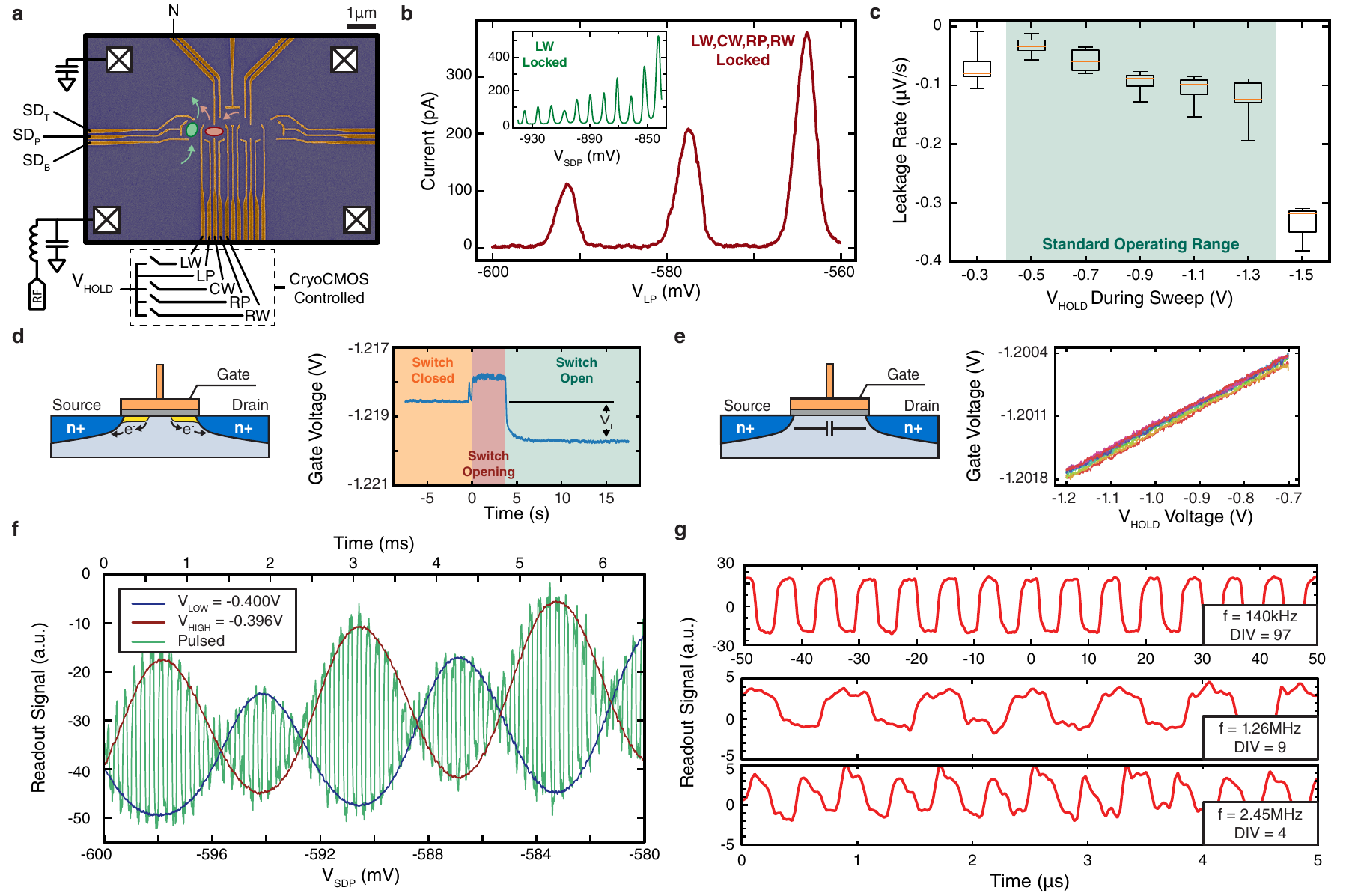}
    \caption{\label{fig:fig3} \textbf{Benchmarking the Cryo-CMOS control with a quantum dot chip.} \textbf{a} False-color SEM of the active area of the quantum dot device. Contacts to the reservoirs are shown schematically as black crosses. Gates used to form quantum dots  are labelled (red and green ovals indicate dots), and current paths used to probe the dots are shown with arrows for each quantum dot. \textbf{b} Single electron tunnelling through the red quantum dot when 4 gates are charge locked by the CLFG cells ($\textrm{V}_\textrm{LP}$ is swept during the measurement). The inset shows a quantum dot (green) where one gate is charge locked and the other gate is varied with a room temperature DAC. \textbf{c} Leakage rate of a voltage held on a gate, after $\textrm{G}_\textrm{LOCK}$ is opened and $\textrm{V}_\textrm{HOLD}$ changed (error bars and statistical analysis are described in Supp. Mat. \textbf{d} Output gate voltage as $\textrm{G}_\textrm{LOCK}$ opened. An offset ($\textrm{V}_\textrm{I}$) between the pre-locked and locked voltage is observed due to charge being injected from the transistor channel to the gate capacitance. The few second delay in the output comes from the large $RC$-time constant of the transport measurement rather than the response of the gate. \textbf{e} Output gate voltage as $\textrm{V}_\textrm{HOLD}$ is swept with the switch open.  A reversible change in locked voltage is observed due to the parasitic source-drain capacitance of the hold switch.  \textbf{f} Increasing the frequency of square wave pulses using an internal frequency divider. \textbf{g} Readout signal through the green quantum dot when the output voltage is referred to $\textrm{V}_\textrm{LOW}$ (blue), $\textrm{V}_\textrm{HIGH}$ (red), and when the switch is rapidly pulsed by the fast gating circuit (green), while the  $\textrm{V}_\textrm{SDP}$ is swept rapidly. The equivalent time is shown along the top axis. For each trace, the voltage on the $\textrm{LW}$ gate is locked at \SI{-1.1}{\volt}. }
\end{figure*}
Our control system generates the large number of static and dynamic voltages needed to manipulate arrays of qubits, without the need for each quantum device to have a direct connection to room temperature electronics. Implemented as a millimeter-scale integrated circuit (IC), the system receives digital instructions via a serial peripheral interface (SPI) requiring just four low-bandwidth wires from room temperature.  Instructions are handled by the digital logic of an on-chip finite-state machine (FSM), which then configures the analog circuit blocks  that interface with qubits. These circuits exploit the low leakage of transistors at cryogenic temperatures to store charge on floating capacitors that include the gate structures needed for biasing and qubit tune-up. By generating static, calibrated voltages in this way, the circuit effectively takes as input a single dc voltage source and multiplexes it into many variable outputs, the number of which, in-principal, can be very large.

Dissipating very little power, this stored charge can then be rapidly shuffled between small capacitors to generate the dynamic voltage pulses required for qubit manipulation\cite{PhysRevB.95.235305,Petta_science05}.  We benchmark this architecture on a GaAs few-electron quantum dot device, characterizing charge leakage and power dissipation under various operating regimes. Based on these results we are then able to project the feasibility of scaling-up this approach. These results demonstrate that complex circuits based on modern CMOS technology can, in fact, be designed to operate near 100 mK, providing a scalable platform for controlling the large number of qubits needed to realise quantum applications. 

\noindent {\bf Experiment}\\
The placement of the CMOS control-chip in relation to the stack of elements needed for quantum computing is shown in Fig. 1a and b. A 30 quantum dot test platform, shown in Fig. 1c, is tightly packaged with the cryo-CMOS chip and also wire bonded to a third chip that comprises superconducting resonators for frequency multiplexed readout \cite{Hornibrook_APL}. The generic platform based on GaAs quantum dots and resonator structures is qubit agnostic, serving as a test vehicle for fast and sensitive measurement of the output signals from the CMOS chip at 100 mK [see Supp. Mat for details]. In order to mitigate unwanted heating of the quantum devices, the chip-packaging arrangement also includes some provision for thermal management by silver pasting each chip to partially-separated gold-plated copper pillars that are in parallel thermal contact to the mixing-chamber stage of a dilution refrigerator [see Fig. 1d].  This arrangement is intended to reduce the direct heat flow from the CMOS circuits to the qubit chip as shown in Fig. 1e. A custom printed circuit board (PCB) and wire-bonds make electrical connection between the chips.

The control-chip is implemented in 28 nm - fully depleted silicon on insulator (FDSOI) technology, a low power and low leakage CMOS platform that is well suited to cryogenic operation\cite{DAS201484,BECKERS2019106,yuanyuan}. Transistors in this FDSOI technology have the utility of configuring a back-gate bias to offset changes in threshold voltage with temperature. This process provides both high (1.8V) and low voltage (1V) devices and also allows for individual back-gate control of n-type and p-type transistors or entire circuit blocks, a useful aspect in mixed-signal circuit design such as our control system.\\ 

\noindent {\bf Control Chip}\\
The floor-plan of our CMOS chip is shown in Fig. 2a and comprises both digital and analog blocks. In the lower left corner of the chip, a series of coupled digital logic circuits provide communication, waveform memory, and autonomous operation of the chip via an FSM (together $\sim$ 100 k transistors). The memory is configured as a 128-bit register, allowing an arbitrary pulse pattern to be stored. A master oscillator is also included, implemented as a ring-oscillator with configurable output frequency. Tiled along the left and bottom edge of the chip is a repeating analog circuit block ``charge-lock fast-gate" (CLFG) that generates the static and dynamic voltages needed for controlling qubits. In the current prototype described here 32 CLFG cells are realized on a single die, enabling connection to 32 qubit-control electrodes [see Supp. Mat. for further details of the circuit blocks].

The basic functionality of a CLFG cell is captured by the circuit shown in Fig. 2b. The voltage $\textrm{V}_\textrm{HOLD}$ is defined by a room temperature digital-to-analog converter (DAC). An individual cell is selected for configuration by the on-chip FSM which connects the output terminal of the cell to the external voltage source by closing the switch $\textrm{G}_\textrm{LOCK,N}$, thereby setting $\textrm{V}_\textrm{OUT,N} = \textrm{V}_\textrm{HOLD}$. This energizes the capacitors required to maintain a static voltage at the high-impedance output. The circuit incorporates an on-chip capacitor $\textrm{C}_\textrm{PULSE}$, part of each CLFG cell, and the parasitic capacitance $\textrm{C}_\textrm{P}$, which includes contributions from the bond-pad, bond-wire and gate-interconnect on the qubit chip. Following charge-up, the switch $\textrm{G}_\textrm{LOCK}$ is opened by the FSM, leaving the charge on the capacitors and qubit gate floating. The low transistor leakage at low temperature allows the charge to be locked for a reasonable time even as the CLFG cell is de-selected, establishing a static voltage at the output. Through sequential cell selection and appropriate adjustment of $\textrm{V}_\textrm{HOLD}$, the single input voltage source can be multiplexed to configure the required voltage biases of many qubit gates. 

For dynamic control, a voltage pulse is required to rapidly change the energy state of the qubit\cite{Petta_science05,gatemon}. Schemes based on MZMs manipulate the qubit via measurement but additionally require large ($\sim$ 100 mV) voltage pulses to open and close tunnel barriers in a specified sequence\cite{PhysRevB.95.235305}. Generating such a pulse with electronics that is significantly decoupled  from the qubit plane requires the use of cables, for instance, that run from the milli-kelvin stage of the refrigerator to 4 K or to room temperature. Driving the impedance of such cables and their attenuators can result in significant power dissipation inside the refrigerator, even if none is dissipated at the end of the open line. Alternatively, by tightly integrating the qubit plane and controller a sizable voltage pulse can be generated with little energy by the redistribution of local charge in a circuit with small capacitance (and high impedance). We exploit this concept in the dynamic operation of the CLFG cell to generate pulses at $\textrm{V}_\textrm{OUT}$. The FSM can be programmed to enable selected cells for pulsing and deliver a pre-loaded pulse pattern to the switch $\textrm{G}_\textrm{FG}$, as illustrated in Fig. 2c. The function of the switch is to toggle the potential of the lower plate of capacitor $\textrm{C}_\textrm{PULSE}$ between two voltage sources $\textrm{V}_\textrm{HIGH}$ and $\textrm{V}_\textrm{LOW}$. These sources can be external to the chip (as they are in our prototype) or derived from local, pre-charged capacitors, as in the static voltage circuit described above. With the potential of the lower plate of $\textrm{C}_\textrm{PULSE}$ switched to $\textrm{V}_\textrm{LOW}$ or $\textrm{V}_\textrm{HIGH}$, charge is induced on the top-plate, changing the output voltage $\textrm{V}_\textrm{OUT}$ that is seen at the qubit gate with respect to ground. The magnitude of the pulse is given by:
\begin{equation}
\nonumber
\Delta \textrm{V}_{\textrm{PULSE}} = \frac{\textrm{C}_{\textrm{PULSE}}}{\textrm{C}_{\textrm{P}} + \textrm{C}_{\textrm{PULSE}}}\left(\textrm{V}_{\textrm{HIGH}} - \textrm{V}_{\textrm{LOW}}\right)
\end{equation}
The power dissipated, $\textrm{P}_\textrm{PULSE}$, is given by the total capacitance, pulse frequency $f$, and voltage of the two levels:
\begin{equation}
\nonumber
\textrm{P}_{\textrm{PULSE}} = \frac{\textrm{C}_{\textrm{PULSE}}    \textrm{C}_{\textrm{P}}   }{\textrm{C}_{\textrm{P}} + \textrm{C}_{\textrm{PULSE}}}\left(\textrm{V}_{\textrm{HIGH}} - \textrm{V}_{\textrm{LOW}}\right)^2 f
\end{equation}
Importantly, as $\textrm{C}_\textrm{P}$ and  $\textrm{C}_\textrm{PULSE}$ are picoFarad capacitances, they require very little power to charge (as determined below). \\

\pagebreak
\noindent{\bf Performance Benchmark}\\
We now turn to benchmark the performance of our CMOS controller using a GaAs-based quantum dot (QD) device [see Supp. Mat for details]. A selection of gates on the QD device are bonded to the output pads of the CLFG cells on the CMOS chip as indicated by the  switched connections drawn at the bottom Fig. 3a. In order to compare the performance of the CMOS circuits to standard control approaches we also connect some of the gates that define the QD to lines directly biased by a room temperature DAC, [see Fig. 3a]. Depending on the combination of gates used, our device\cite{Croot} can be configured to form quantum dots in various locations as indicated by the red and green ovals in Fig. 3a. As a basic demonstration of multiplexing through charge-locking, we program four CLFG cells to bias four gates, using a single external voltage source, to create the quantum dot shown in red in Fig. 3a. Transport current through this dot, as a function of $\textrm{V}_\textrm{LP}$ (also routed through a 5th CLFG block) exhibits familiar Coulomb blockade oscillations, as shown in Fig. 3b. 

A second quantum dot, shown in green in Fig. 3a, is also configured in the same device to measure leakage of CLFG cells, with transport data shown in the inset of Fig. 3b for the case where $\textrm{V}_\textrm{LW}$ is locked and $\textrm{V}_\textrm{SDP}$ is swept (see Fig. 3a for gate labels). Biasing this dot to the edge of a Coulomb blockade peak and measuring the transport current as a function of time allows charge leakage from the CLFG cell to be directly detected. The charge leakage leads to the voltage on the gate changing (deterministically) at a rate of order 1 part in 10$^7$ per second (10s of $\mu$V / Hr) and is dependent on the value of $\textrm{V}_\textrm{HOLD}$, as shown in Fig. 3c. This leakage rate is sufficiently low to enable a ``round-robin" refresh cycle every few minutes to lock and stablize many gates with a single DAC input to the control chip. Decreasing the leakage further is possible by increasing the capacitance of the CLFG cell, at the expense of the pulse rise-time and chip footprint. 

Measurements of the QD conductance, shown in Fig. 3d, also provide a means  of monitoring the charge-locking process, including charge injection at the CMOS switch, which pushes additional charge onto the gate capacitance as the switch is opened [see cartoon in Fig. 3d and Supp. Mat.]. There is also a linear offset in the gate voltage that depends on the value of $\textrm{V}_\textrm{HOLD}$, as shown in Fig. 3e. This offset comes from the charge induced on the gate via the transistor capacitance when the switch is open [see cartoon in Fig. 3e]. In our setup both charge injection and linear offset effects are calibrated and transparently accounted for in software that interfaces with the control chip.

Our quantum dot structure can also be configured as an rf single electron transistor (rf-SET) by embedding it in an impedance matching $LC$ tank circuit\cite{Reilly:2007ig}, an aspect that allows detection of the gate-pulsing action of the CMOS chip with a bandwidth of $\sim$ 10 MHz. As a first demonstration, we program the cryo-CMOS chip to charge-lock the gate-electrodes and switch $\textrm{G}_\textrm{FG}$ at 140 kHz. The CLFG voltage pulse amplitude is calibrated to switch the dot conductance from the top of a Coulomb peak to a trough. Simultaneously sweeping a second gate voltage ($\textrm{V}_\textrm{SDP}$), we observe the modulation of dot conductance whose envelope maps out Coulomb blockade oscillations, as shown by the green trace in Fig. 3f. The envelope overlaps with the red and blue traces in the figure, which correspond to the direct measurement of the oscillations by setting $\textrm{V}_\textrm{HIGH}$ or  $\textrm{V}_\textrm{LOW}$ and sweeping $\textrm{V}_\textrm{SDP}$ without pulsing, thus verifying the action of the CMOS pulsing circuit. Disabling the $\textrm{V}_\textrm{SDP}$ sweep, we bias the device to directly observe the conductance modulation as a continuous square wave. Configuring the CMOS oscillator circuit to modify the frequency of the pulses, as shown in Fig. 3g. Note that the apparent rise-time of these pulses is limited by the bandwidth of the $LC$ tank circuit. The true rise-time, set by the $RC$ constant of CMOS switches and total capacitance, is of order a few nanoseconds. Beyond verifying the action of the CMOS pulsing circuit, these results demonstrate the suitability of the controller for manipulating single electron states by rapidly varying the chemical potential of a device as required for qubit control.

We also measure the temperature of the packaged system as each circuit block in our control-chip is powered up sequentially, as shown in Fig. 4a. Separating the contributions arising from the clock generator, FSM, and CLFG cells (including $\textrm{C}_\textrm{PULSE}$ and $\textrm{C}_\textrm{P}$), the data in Fig. 4a shows that the temperature remains below 100 mK for frequencies of order a few MHz. In our setup the temperature is measured by a thermometer in close proximity to the qubit chip and is significantly higher  than the temperature measured by a second thermometer at the mixing-chamber stage of the refrigerator (96 mK compared to 36 mK when running at 5.1 MHz). The presence of such a thermal gradient suggests that lower temperatures are possible with improved packaging that achieves better thermalization of the chips to the refrigerator.

\begin{figure*}
\includegraphics[width=\textwidth]{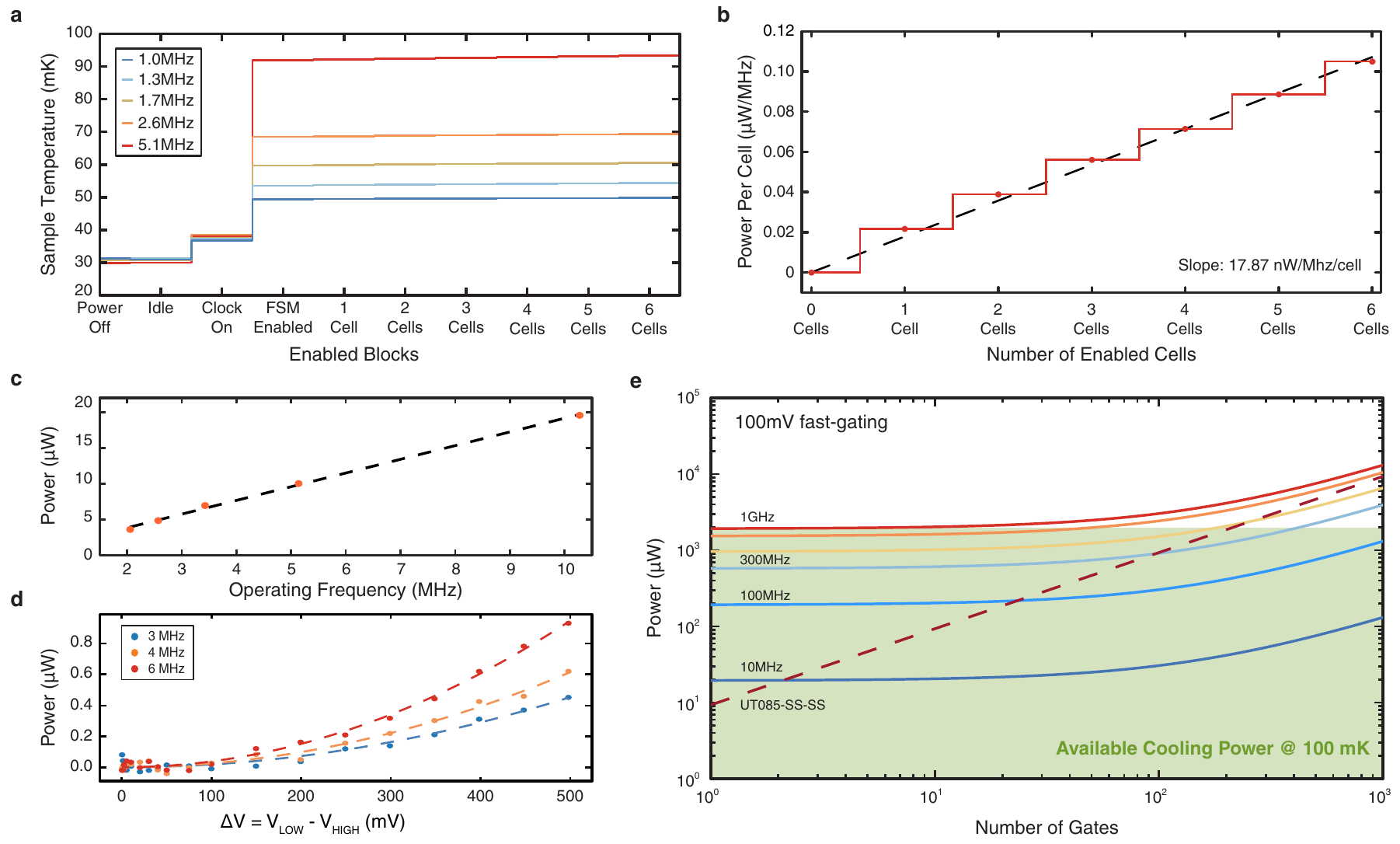}
\caption{\label{fig:fig4} \textbf{Power dissipated at 100 mK when interfacing with quantum dot device.} \textbf{a} Temperature of the system for various operating modes of the Cryo-CMOS chip. The steady state temperature progressively increases as the on-chip clock and FSM block is enabled, followed by simultaneous fast pulsing on 1 to 6 output gates in sequence. \textbf{b} Calibrated incremental power generated by Cryo-CMOS  while applying 0.1 V pulsing on each qubit gate using CLFG cells 1 -6. We extract the average heat cost to be 18 nW / MHz per cell. Each sub-system in the Cryo-CMOS control IC is measured in isolation and its heat load is characterised.  \textbf{c}  Power dissipated by the FSM block for different operating frequencies, and \textbf{d} plots the measured $C \Delta V^2f$ power for different fast-gating frequencies and amplitudes. \textbf{e} Projected total system power of the Cryo-CMOS controller as a function of the number of qubit gates and operating frequency, based on the measured power dissipation. Shaded green region indicates the cooling power that is achievable with today's commercial dilution refrigerators, keeping the qubits at 100 mK.}
\end{figure*}
These temperature measurements can then be used to indicate the power dissipated by each CMOS circuit block. To measure the dissipation we first record the temperature of the thermometer as a function of heat applied to a resistor inside the CMOS chip with a known resistance and measured voltage bias. This provides a conversion between on-chip power dissipation and temperature [see Supp. Mat. for details].  For the purpose of extrapolating our results to large numbers of control lines we determine the power of each block, showing the independent contribution from the  CLFG cells in Fig. 4b and FSM in Fig. 4c. We observe the expected quadratic dependence of the power with gate voltage amplitude, as shown in Fig. 4d. \\

\noindent{\bf Discussion}\\
Combining these results, the total system power can then be determined as a function of frequency and the number of output gates. Setting the pulse amplitude at 100 mV (needed for controlling MZM qubits), Fig. 4e shows the total system power. The green-shaded region indicates the cooling power provided by commercially available dilution refrigerators\cite{leiden}. As a comparison we also plot the effective power dissipated in a standard coaxial transmission line of the kind commonly used for qubit control (UT-085-SS-SS with 3 dB attenuation) \cite{WallraffPRApp}.   Given that for qubits based on MZMs \cite{PhysRevB.95.235305} the clock speed of quantum logic gates is a few MHz, it appears from Fig. 4e that CMOS-based control is a viable approach for thousands of qubits.

Controlling large numbers of alternative qubit platforms based on electron spins or gatemons\cite{gatemon} also appears within reach using this CMOS-based approach. In such systems the CMOS circuits can provide either direct control \cite{Petta_science05} or be configured to use the output of CLFG cells to pulse a gate that brings the qubit into resonance with a microwave tone. Further, the circuits described here can be combined with a cryogenic switch matrix\cite{PhysRevAppliedJohn}, steering microwave pulses to the appropriate qubit under the control of our CMOS platform. Somewhat unconstrained by power dissipation at the 1000-qubit level, it is worth also considering the footprint of CLFG cells in a scaled-up implementation. Given a cell size of $\sim$ 100 $\mu$m$^2$, as we have in our prototype chip, a 1000-cell implementation occupies an area of a few mm$^2$. Even at  gate counts in the millions, the size of the control-chip ($\sim$ 10 cm$^2$) is within what is possible today with conventional CMOS fabrication. \\

\noindent{\bf Acknowledgements}\\
This research was supported by Microsoft Corporation and the ARC Centre of Excellence for Engineered Quantum Systems. We thank R. Rouse for help with chip tape-out, M. Cassidy, S. Waddy, C. Marcus, and L. Kouwenhoven  for discussions. We acknowledge the facilities as well as the scientific and technical assistance of the Research \& Prototype Foundry Core Research Facility at the University of Sydney, part of the Australian National Fabrication Facility (ANFF).\\
$\dagger$ Corresponding author: david.reilly@sydney.edu.au\\

%

\end{document}